# A FORCE-FREE MAGNETIC FIELD SOLUTION IN CYCLIDAL COORDINATES

**Gerald E. Marsh**
Argonne National Laboratory (Ret)
gemarsh@uchicago.edu
Chicago, Illinois, 60615, USA

## ABSTRACT

There are no known analytic solutions to the force-free magnetic field equations in coordinates representing Dupin cyclides. The fact that the curvature lines of Dupin cyclides are all circles is used here to derive a force-free magnetic field on the surface of an elliptic Dupin cyclide.

# INTRODUCTION

In plasma physics, where the condition for plasma equilibrium is given by $(\nabla \times \boldsymbol{B}) \times \boldsymbol{B} = \nabla p$, where $p$ is the plasma pressure, the magnetic field will be force-free if $\nabla p = 0$. Force-free means that the "self-force" or Lorentz force vanishes. Force-free magnetic field configurations are difficult to find because $(\nabla \times \boldsymbol{B}) \times \boldsymbol{B} = 0$ is a nonlinear equation. The plasma $\beta$ is defined as the ratio of the plasma pressure to the magnetic pressure $p_m$. The force-free approximation is valid for "low-beta" plasmas. Such plasmas are often found in an astrophysical context. Here it is shown that force-free fields can not only have solutions for toroids, as was shown in an earlier paper, but also for cyclides.

Dupin cyclides are those surfaces where the curvature lines are circles. The curvature lines are given by the parametric variables $u$ and $v$ with one or the other being equal to a constant. Of course, this is also true for tori, which—as will be discussed below—arise from setting two of the parameters of an elliptic cyclide equal. An elliptic cyclide has a parametric representation in three-dimensional Cartesian coordinates as[1]

$$x = \frac{d(c - a\,cosu\,cosv) + b^2 cosu}{a - c\,cosu\,cosv},$$

$$y = \frac{b\,sinu(a - dcosv)}{a - c\,cosu\,cosv},$$

$$z = \frac{b\,sinv(c\,cosu - d)}{a - c\,cosu\,cosv},$$

$$0 \leq u, v < 2\pi.$$

(Eqs. 1)

Figure 1 gives an example of an elliptic cyclide and also a cutaway view of nested elliptic cyclides. The lines of curvature are given by $u$ and $v$ with one or the other being equal to a constant; i.e., for a constant value of $v$, $u$ follows the curvature circle around the cyclide the long way and for a constant value of $u$, $v$ follows the curvature circle around the cyclide the short way.

Focal conics can be seen as degenerate focal surfaces: Dupin cyclides are the only surfaces where focal surfaces collapse to a pair of curves that are focal conics. In the case of the elliptic cyclides the focal conics are an ellipse and a hyperbola. The plane of the hyperbola is orthogonal to the plane containing the ellipse and the hyperbola is the focal conic to the ellipse.

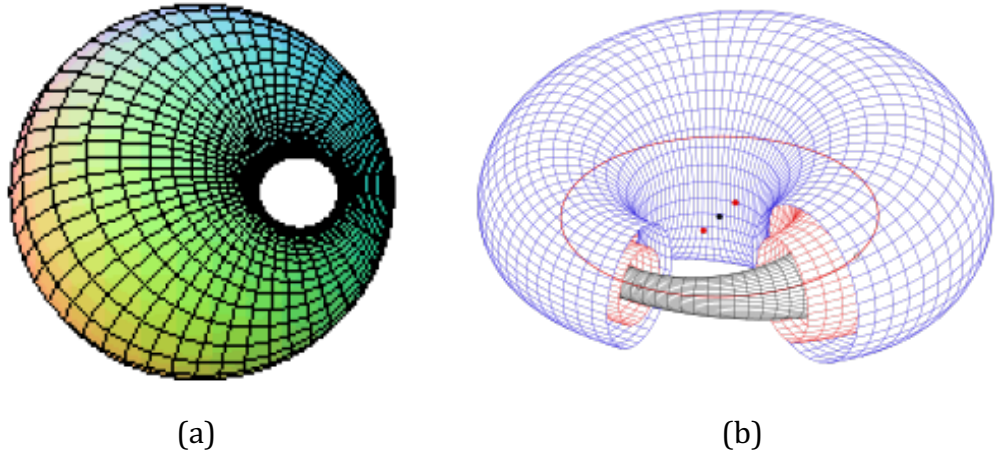

(a) (b)

Figure 1. (a) An elliptic cyclide. Note the perpendicular lines of curvature. (b) Parallel or nested surfaces of an elliptic cyclide. The parameters of the surfaces are: $a = 1$, $b = 0.98$, $c = 0.199$ and d, which specifies each of the cyclide surfaces, is $d = 0.30, 0.45$, and $0.60$. The ellipse shown in the figure is discussed below [Source: Wikimedia Commons].

The parameters $a$, $b$, $c$, $d$ in Eqs. (1) are the semi-major and semi-minor axes of the ellipse and $c$ is its eccentricity. A Dupin cyclide can be considered to be a channel surface (the envelope of a one-parameter family of spheres) and $d$ would be the average radius of the generating spheres. The equations of the ellipse and hyperbola are

$$\frac{x^2}{a^2} + \frac{y^2}{b^2} = 1, \quad z = 0,$$
$$\frac{x^2}{c^2} - \frac{z^2}{b^2} = 1, \quad y = 0,$$
$$0 < b \leq a, \; c^2 = a^2 - b^2.$$

Eqs. (2)

The implicit equation,[2] meaning that the equation is not solved for the variables $x, y$, or $z$, is important for relating elliptic cyclides to tori of revolution. It is given by

$$(x^2 + y^2 + z^2 + b^2 - d^2)^2 - 4(ax - cd)^2 - 4b^2y^2 = 0.$$

Eq. (3)

In Eq. (3) some authors omit the factor 4 in the last term. The idea is to set $a = b$ and show that $c = 0$ so that the ellipse becomes a circle and the hyperbola degenerates into a line so that the corresponding degenerate cyclides are tori of revolution thus showing that the torus belongs to the class of elliptic cyclides. However, Eq. (3) is very difficult to solve for $c$ because the coefficients are inexact, but if one sets $a = b = 1$ and $d = 0.3$, one can obtain a solution by solving a corresponding exact system and "numericizing" the result. It turns out that $c$ is indeed equal to zero.

### THE FORCE-FREE MAGNETIC FIELD EQUATIONS IN CYCLIDAL COORDINATES

The general solution to the force-free field equations in cyclidal coordinates is a very difficult problem. The force-free field equation is $\nabla \times \boldsymbol{B} = \alpha \boldsymbol{B}$, where the expressions for the curl or divergence are non-illuminating complex functions of trigonometric functions that are very, very long. Some simplifications must be imposed. Because the divergence of the magnetic field must vanish, it is assumed here that the normal component of the magnetic field must vanish on the surface of the cyclide. Strictly speaking, a cyclide does not have cylindrical symmetry, making finding a solution almost impossible.

But, as discussed above, the lines of curvature of any Dupin cyclide are circles having the parametric coordinates $u$ and $v$. To obtain a viable solution to the force-free magnetic field equations, cylindrical symmetry is assumed for each of the $u$ circular coordinate lines upon which $v$ is a constant. This means that $B_u$ is constant along each $u$-circle. This is unusual because the planes containing the $u$-circles where $v$ equals a constant are not parallel.

If $w$ represents the coordinate normal to the cyclidal surface, the metric coefficients in cyclidal coordinates, which can be found from Eqs. (1) are:

$$h_u = \sqrt{\left(\frac{1}{(a - c\cos(u)\cos(v))^4}\right.}$$
$$\left(\sin^2(u)\left(a\,b^2 - d\left(a^2 - c^2\right)\cos(v)\right)^2 + \right.$$
$$b^2\,c^2\sin^2(u)\sin^2(v)\,(a - d\cos(v))^2 +$$
$$\left.\left.b^2\,(a - d\cos(v))^2\,(a\cos(u) - c\cos(v))^2\right)\right),$$

$$h_v = \sqrt{\left(\frac{1}{(a - c\cos(u)\cos(v))^4}\right.}$$
$$\left(\cos^2(u)\sin^2(v)\left(d\left(c^2 - a^2\right) + b^2\,c\cos(u)\right)^2 + \right.$$
$$a^2\,b^2\sin^2(u)\sin^2(v)\,(d - c\cos(u))^2 +$$
$$\left.\left.b^2\,(d - c\cos(u))^2\,(c\cos(u) - a\cos(v))^2\right)\right),$$

$$h_w = \left(\left(a^2\cos^2(u)\cos^2(v) - 2\,a\,c\cos(u)\cos(v) + b^2\sin^2(u)\cos^2(v) + \right.\right.$$
$$\left.\left.b^2\sin^2(v) + c^2\right) \Big/ (a - c\cos(u)\cos(v))^2\right).$$

Eqs. (4)

In arbitrary curvilinear coordinates the divergence of $\boldsymbol{B}$ is given by

$$\nabla \cdot \boldsymbol{B} = \frac{1}{\sqrt{g}} \partial_{x_i}(\sqrt{g} B^i),$$

Eq. (5)

where $g = |g_{ij}|$, $i = u, v, w$. The calculations for the force-free fields will be greatly simplified if advantage is taken of the requirement that the normal component of the magnetic field, $B_w$, must vanish on the surface of the cyclide. This allows the introduction of a flux function $\Phi$. Equation (5) will then be satisfied if a function is introduced such that

$$\sqrt{g} B^u = \partial_v \Phi, \qquad \sqrt{g} B^v = -\partial_u \Phi.$$

Eq. (6)

A Dupin cyclide has orthogonal coordinates, and for such coordinates $g_{ij} = g^{ij} = 0, i \neq j$; the metric coefficients $h_i$ are defined by $g_{ii} = h_i^2$, $g^{ii} = 1/h_i^2$; $\sqrt{g} = h_u h_v h_w$. And the physical components are then $h_i B^i$. From Eq. (6) these may be written as

$$B_u = \frac{1}{h_v h_w} \partial_v \Phi, \qquad B_v = -\frac{1}{h_u h_w} \partial_u \Phi.$$

Eqs. (7)

What remains to do in order to use this approach of making use of a flux function is to find the function itself and show that the fields of Eq. (7) are solutions of the force-free field equations. However, finding the flux function in coordinates far simpler than cyclidal coordinates can be a very difficult problem.[3] There is, however, a simpler way to find the flux function: Equations (7) imply that

$$d\Phi = h_v h_w B_u - h_u h_w B_v,$$

Eq. (8)

so that given test functions for $B_u$ and $B_v$ one can integrate Eq. (8) to find $\Phi$. Finding appropriate test functions is a question of intuition. What is done here is to choose one that looks interesting when mapped onto the cyclide. Here is a simple example that will used

$$B_u = 2/h_v \text{ and } B_v = 10/h_u.$$

Eqs. (9)

The flux function is then

$$\Phi = 2 \int h_w dv - 10 \int h_w du.$$

Eq. (10)

Given the coefficients $a, b, c, d$, the flux function $\Phi$ is:

$$\Phi = \frac{1}{c^2}\left(5\left(\frac{8\sqrt{2}\,a\,(a^2-b^2-c^2)\,\text{ArcTanh}\left[\frac{\sqrt{2}\,(a+c\cos[v])\tan\left[\frac{u}{2}\right]}{\sqrt{-2a^2+c^2+c^2\cos[2v]}}\right](a^2-c^2\cos[2v])}{(-2a^2+c^2+c^2\cos[2v])^{3/2}}\right.\right.$$

$$+ \left(-2a(a^2-b^2)c^2u\cos[v]^2 + a(a^2-b^2)u(4a^2-c^2-c^2\cos[2v]) \right.$$
$$+ 2(a^2-b^2)cu\cos[u]\cos[v](-2a^2+c^2+c^2\cos[2v])$$
$$\left.+ 4c(-a^2+c^2)(-a^2+b^2+c^2)\cos[v]\sin[u]\right) \Big/$$

$$\left.\left.\Big((a-c\cos[u]\cos[v])(-2a^2+c^2+c^2\cos[2v])\Big)\right)\right) -$$

$$\frac{1}{c^2}\left(\frac{8\sqrt{2}\,a\,(a^2-b^2-c^2)\,\text{ArcTanh}\left[\frac{\sqrt{2}\,(a+c\cos[u])\tan\left[\frac{v}{2}\right]}{\sqrt{-2a^2+c^2+c^2\cos[2u]}}\right](a^2-c^2\cos[2u])}{(-2a^2+c^2+c^2\cos[2u])^{3/2}}\right.$$

$$+ \left(-2a(a^2-b^2)c^2v\cos[u]^2 + a(a^2-b^2)v(4a^2-c^2-c^2\cos[2u]) \right.$$
$$+ (a^2-b^2)c^3v\cos[u]^3\cos[v]$$
$$+ c\cos[u]\left(-\frac{1}{2}(a^2-b^2)v(8a^2-3c^2-3c^2\cos[2u])\cos[v]\right.$$
$$\left.\left.+ 4(a^2-c^2)(a^2-b^2-c^2)\sin[v]\right)\right) \Big/$$

$$\left.\Big((-2a^2+c^2+c^2\cos[2u])(a-c\cos[u]\cos[v])\Big)\right).$$

Eq. (11)

Putting in the numerical value for the constants of *a* = 1, *b* = 0.98, *c* = 0.199, and *d* = 0.3, one can compute the flux function. The stream and vector plots of the fields in Eqs. (9) are shown below and then mapped onto the cyclide:

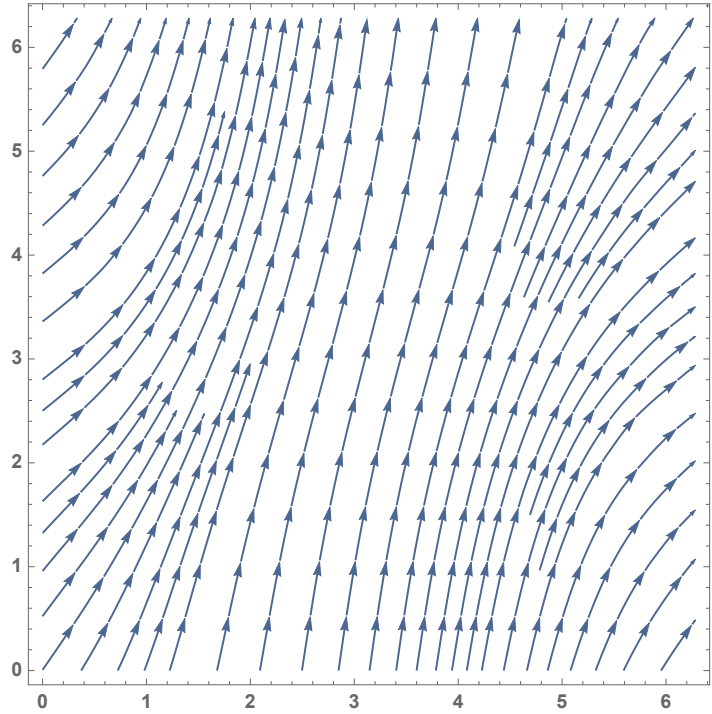

Figure 2. Stream plot of the magnetic field components given by Eqs. (9). The *u*-axis is along the abscissa and the *v*-coordinate along the ordinate.

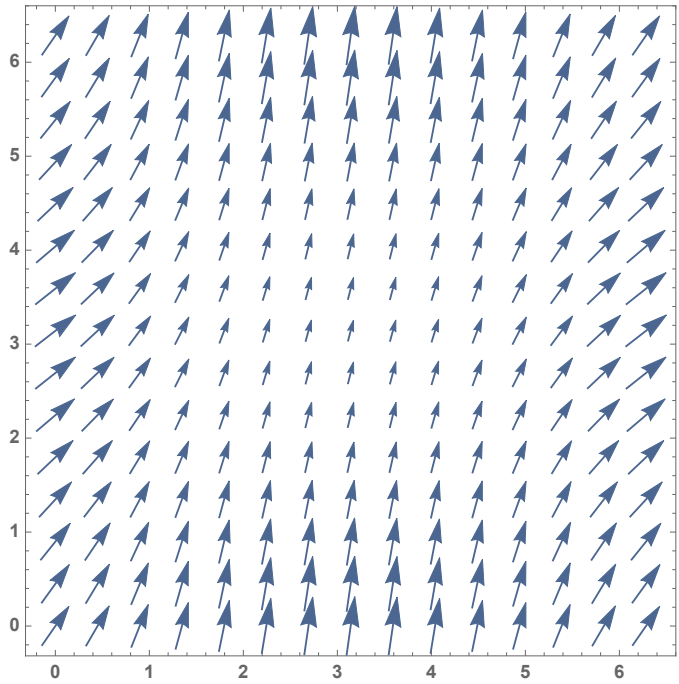

Figure 3. Vector plot of the magnetic field components given by Eqs. (9). The *u*-axis is again along the abscissa and the *v*-coordinate along the ordinate.

The stream plot mapped onto the cyclide, using the numerical values of the constants given above, is shown in Fig. 4, and the vector plot on the cyclide by Fig. 5.

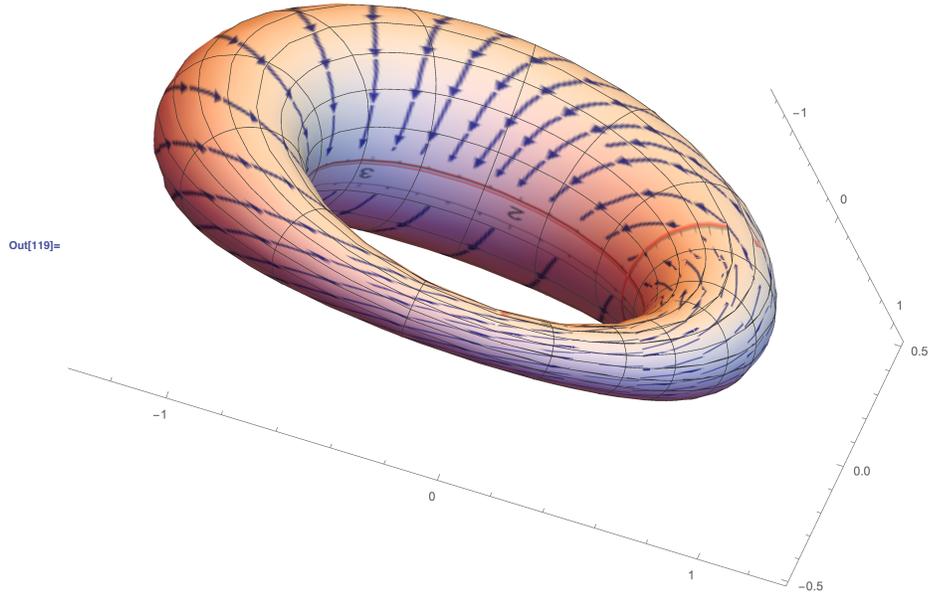

Figure 4. Stream plot of the magnetic field components specified by the flux function Φ, given by Eq. (11), mapped onto the cyclidal surface.

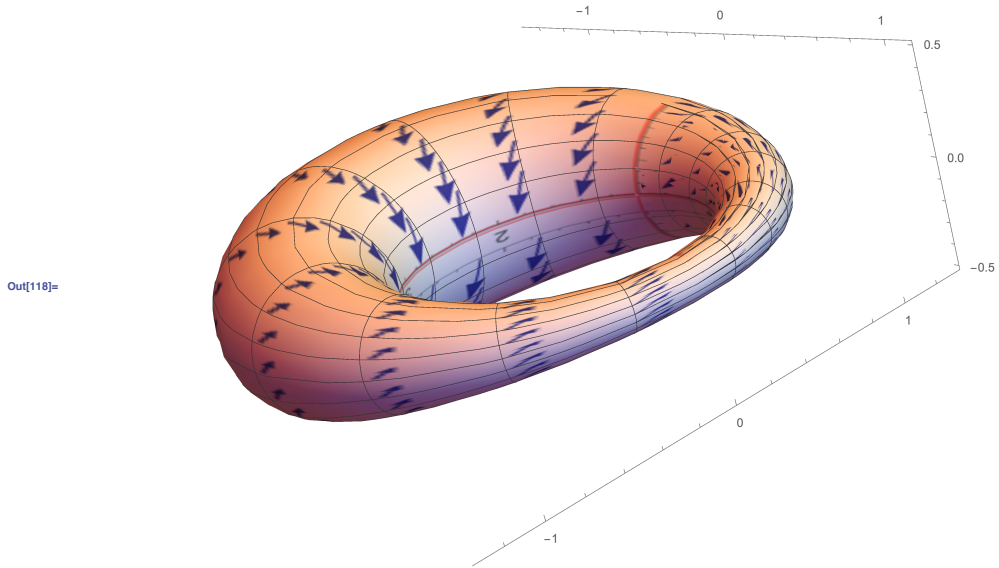

Figure 5. Vector plot of the magnetic field components specified by the flux function Φ, given by Eq. (11), mapped onto the cyclidal surface.

It remains to show that the force-free field equations are satisfied by the magnetic fields of Eqs. (9) by finding an expression for the associated function $\alpha$.

The force-free field equation in 3-dimensional space is given by $\nabla \times \boldsymbol{B} = \alpha \boldsymbol{B}$; but it has been assumed here that the normal component of the magnetic field must vanish on the surface of the cyclide so that the divergence of the magnetic field vanishes. This means that the field equation must be evaluated on the surface of the cyclide and the 3-dimensional curl cannot be used.

Consequently, on the surface one must use the 2-dimensional component of the curl, also known as the circulation in another context. It is given by

$$(\nabla \times \boldsymbol{B})_w = \frac{1}{h_u h_v}(\partial_u (h_v B_v) - \partial_v (h_u B_u)).$$

Eq. (12)

The full equation for $(\nabla \times \boldsymbol{B})_w$ with the meanings of the symbols inserted is very long and consequently is shown in the Appendix.

The force-free field equation to be solved for the function $\alpha$ is then

$$\frac{1}{h_u h_v}(\partial_u (h_v B_v) - \partial_v (h_u B_u)) = \alpha (B_u + B_v).$$

Eq. (13)

The solution for $\alpha$ is again very long and is also shown in the Appendix.

Figures 6(a) and 6(b) show 3-dimensional plots of the function $\alpha$. The cross sections at $u = 0$ and $u = 2\pi$ match since they represent the same cross section on the cyclide and the same is true for $v = 0$ and $v = 2\pi$. Note the changes in sign of the function $\alpha$. In the regions where $\alpha$ is zero, the curl vanishes and the magnetic field is the gradient of a scalar function.

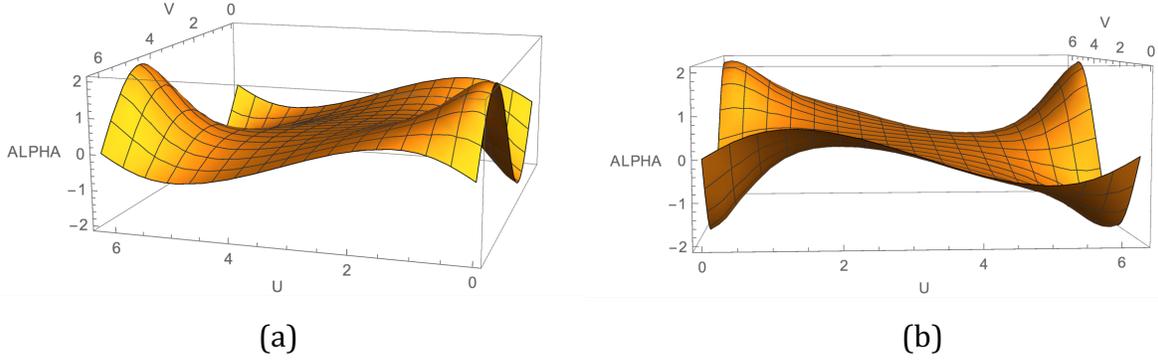

(a)  (b)

Figure 6.  3-dimensional plots of the solution for the function $\alpha$ of Eq. (13).  The variables $u$ and $v$ are shown as capitals in the figures to make them easier to read.

The solution given above is then an example of a solution to the problem of finding a force-free magnetic field on the surface of a cyclide.

## CYCLIDE TO TOROID

As mentioned in the Introduction, the torus belongs to the class of elliptic cyclides.  By setting the coefficients $a = b$ and $c = 0$, the flux function approach to finding solutions used above should yield a solution on a torus.  It is shown in this section that that is indeed true.  To make comparison easier the values of the constants are now chosen to be $a = b = 1$, $c = 0$, and $d = 0.3$.  With these values the metric coefficients of Eqs. (4) become

$$h_u = \sqrt{\cos^2(u)(1 - 0.3\cos(v))^2 + \sin^2(u)(1 - 0.3\cos(v))^2},$$
$$h_v = \sqrt{0.09\sin^2(u)\sin^2(v) + 0.09\cos^2(u)\sin^2(v) + 0.09\cos^2(v)},$$
$$h_w = \cos^2(u)\cos^2(v) + \sin^2(u)\cos^2(v) + \sin^2(v).$$

Eqs. (14)

And, using the same magnet field components as used above, we obtain for the flux function the simple result $\Phi = -10u + 2v$.

The stream and vector plots for the degenerate cyclide are shown in Fig. (7)

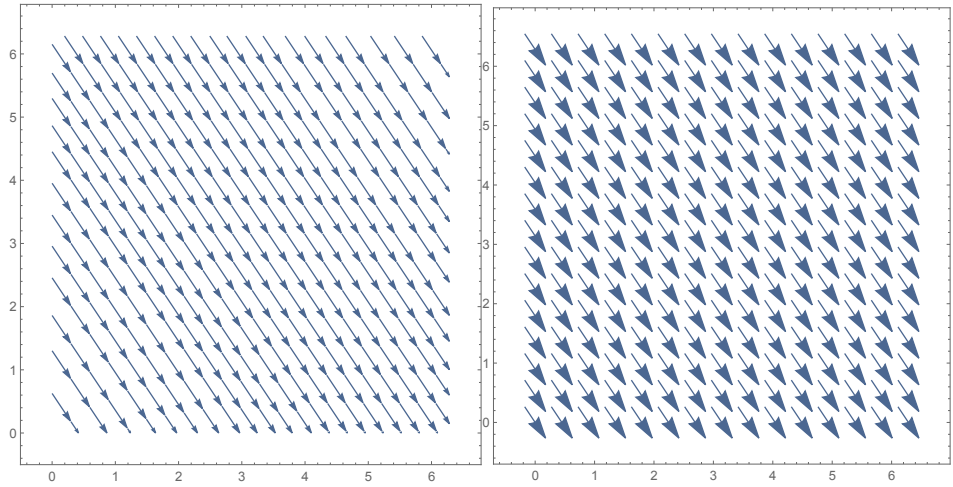

Figure 7. Stream and vector plots of the magnetic field components given by Eqs. (9), but in this case for the degenerate cyclide. The *u*-axis is again along the abscissa and the *v*-coordinate along the ordinate.

When the stream and vector plots are mapped onto the degenerate cyclide, as shown in Figs. (8), it is obvious that the degenerate cyclide is indeed a torus.

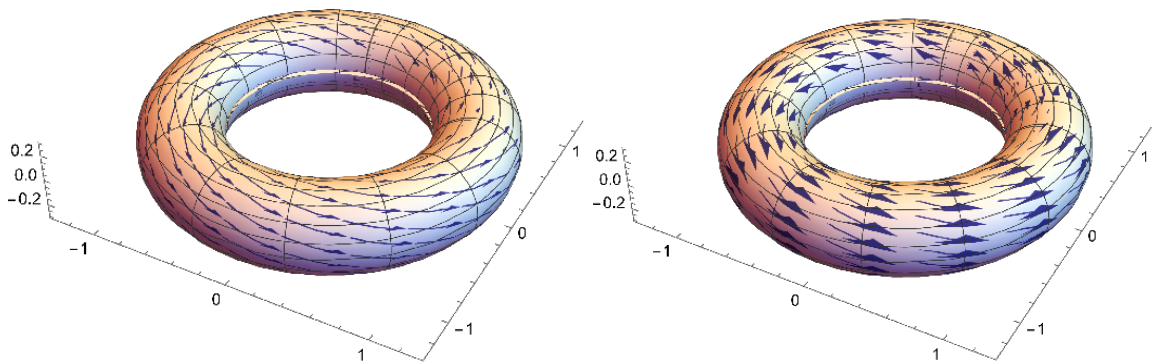

Figure 8. Stream and vector plots of the magnetic field components given by Eqs. (9), mapped onto the degenerate cyclide.

The plot of the magnetic field mapped on the torus is different and much less complex than that found in Reference 4. The flux function approach can be used to find many other force-free magnetic field solutions for the torus.

In order to show that this example is a solution to the force-free field equations, it must be demonstrated that there exists an appropriate function $\alpha$. Solving for the function $\alpha$ using Eq. (13) results in

$$\alpha = \left(1.` \left(-\left(\left(1.` \left(0.6` \cos^2(u) \sin(v) (1.` - 0.3` \cos(v)) + 0.6` \sin^2(u) \sin(v) (1.` - 0.3` \cos(v))\right)\right) \Big/ \left(\sqrt{0.09` \sin^2(u) \sin^2(v) + 0.09` \cos^2(u) \sin^2(v) + 0.09` \cos^2(v)} \sqrt{\cos^2(u) (1.` - 0.3` \cos(v))^2 + \sin^2(u) (1.` - 0.3` \cos(v))^2}\right)\right) + \left(\left(0.18` \cos^2(u) \sin(v) \cos(v) + 0.18` \sin^2(u) \sin(v) \cos(v) - 0.18` \sin(v) \cos(v)\right) \sqrt{\cos^2(u) (1.` - 0.3` \cos(v))^2 + \sin^2(u) (1.` - 0.3` \cos(v))^2}\right) \Big/ \left(0.09` \sin^2(u) \sin^2(v) + 0.09` \cos^2(u) \sin^2(v) + 0.09` \cos^2(v)\right)^{3/2} + 0.`\right)\right) \Big/ \left(\sqrt{0.09` \sin^2(u) \sin^2(v) + 0.09` \cos^2(u) \sin^2(v) + 0.09` \cos^2(v)} \left(\frac{2.`}{\sqrt{0.09` \sin^2(u) \sin^2(v) + 0.09` \cos^2(u) \sin^2(v) + 0.09` \cos^2(v)}} - \frac{10.`}{\cos^2(u) \cos^2(v) + \sin^2(u) \cos^2(v) + \sin^2(v)}\right) \sqrt{\cos^2(u) (1.` - 0.3` \cos(v))^2 + \sin^2(u) (1.` - 0.3` \cos(v))^2}\right)$$

Eq. (15)

Here is a 3-dimensional plot of the function $\alpha$

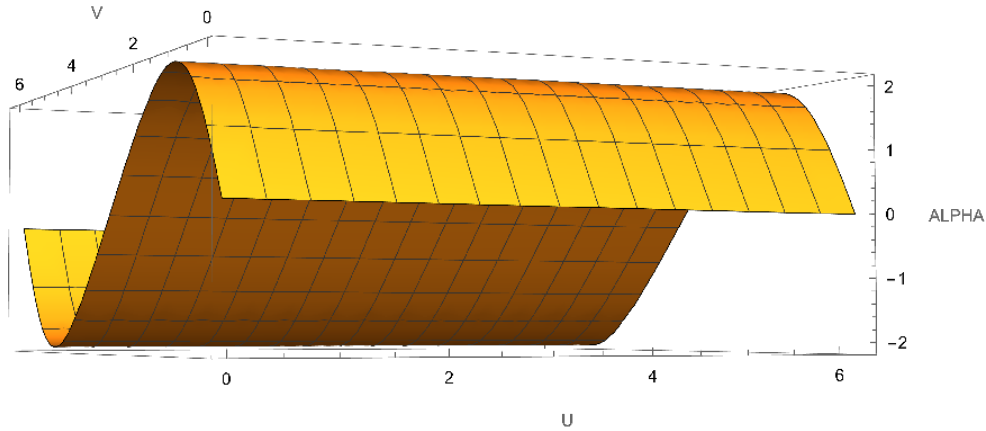

Figure 9. A 3-dimensional plot of the function $\alpha$ for the degenerate cyclide/torus.

The cross sections at $u = 0$ and $u = 2\pi$ match since they represent the same cross section on the torus and the same is true for $v = 0$ and $v = 2\pi$. Note the changes in sign of the function $\alpha$. Where $\alpha$ is zero, for the degenerate cyclide/torus at $v = 0, \pi$, and $2\pi$, the curl vanishes and the magnetic field is the gradient of a scalar function.

## SUMMARY

Recently, the problem of finding force-free magnetic fields on a torus has been solved[4], and since the torus can be viewed as a degenerate cyclide the question naturally arises as to whether the more general problem of finding a force-free magnetic field on the surface of a cyclide can be solved for a Dupin cyclide. The torus has cylindrical symmetry which greatly simplifies the problem, but this is not the case for a cyclide. However, the lines of curvature of any Dupin cyclide are circles, allowing the introduction of a limited form of cylindrical symmetry for each of the $u$ circular coordinate lines, upon which $v$ is a constant. This results in a greatly simplified problem for finding a solution for a magnetic field on the cyclide, but it should be remembered that the planes containing the $u$-circles where v is constant are not parallel.

If the magnetic field is independent of one coordinate, chosen above to be the normal to the surface of the cyclide, the divergence of the magnetic field will vanish if a flux function — see Eq. (6)—is introduced. Since the coordinates introduced for the cyclide are orthogonal, the components of the magnetic field may be written as in Eq. (7). By using test functions for $B_u$ and $B_v$ one can find $\Phi$. The previous section explains the choice of the simple test functions specified in Eqs. (9).

Since it was assumed that the normal component of the magnetic field vanishes on the surface of the cyclide so as to have the divergence of the magnetic field vanish, the 3-dimensional form of the curl cannot be used in the force-free field equation. Instead, one must use the 2-dimensional component of the curl to obtain Eq. (13) which is then solved for the function $\alpha$. Figure (6) shows that this function has negative values for some regions on the surface of the cyclide. In the regions where $\alpha$ is zero, the curl vanishes and the magnetic field is the gradient of a scalar function.

# APPENDIX

The equations shown below are not particularly enlightening and are only given here for completeness. The font size is limited by the transfer from Mathematica to MS Word. The 2-dimensional curl is:

$(\nabla \times \boldsymbol{B})_w =$

$$\Bigg(\bigg(\sqrt{\Big(\tfrac{1}{(a-c\cos(u)\cos(v))^4}(b^2(a\cos(u)-c\cos(v))^2(a-d\cos(v))^2+b^2c^2\sin^2(u)\sin^2(v)(a-d\cos(v))^2+(ab^2-(a^2-c^2)d\cos(v))^2\sin^2(u))\Big)}$$
$$\Big(\tfrac{1}{(a-c\cos(u)\cos(v))^4}(2a^2b^2\cos(v)\sin^2(u)\sin(v)(d-c\cos(u))^2+2ab^2(c\cos(u)-a\cos(v))\sin(v)(d-c\cos(u))^2+$$
$$2\cos^2(u)(c\cos(u)b^2+(c^2-a^2)d)^2\cos(v)\sin(v))-\tfrac{1}{(a-c\cos(u)\cos(v))^5}4c\cos(u)\sin(v)$$
$$(b^2(d-c\cos(u))^2(c\cos(u)-a\cos(v))^2+\cos^2(u)(c\cos(u)b^2+(c^2-a^2)d)^2\sin^2(v)+a^2b^2(d-c\cos(u))^2\sin^2(u)\sin^2(v))\Big)\bigg/$$
$$\Big(\tfrac{1}{(a-c\cos(u)\cos(v))^4}(b^2(d-c\cos(u))^2(c\cos(u)-a\cos(v))^2+\cos^2(u)(c\cos(u)b^2+(c^2-a^2)d)^2\sin^2(v)+$$
$$a^2b^2(d-c\cos(u))^2\sin^2(u)\sin^2(v))\Big)^{3/2}-$$
$$\Big(5\sqrt{\Big(\tfrac{1}{(a-c\cos(u)\cos(v))^4}(b^2(d-c\cos(u))^2(c\cos(u)-a\cos(v))^2+\cos^2(u)(c\cos(u)b^2+(c^2-a^2)d)^2\sin^2(v)+}$$
$$\overline{a^2b^2(d-c\cos(u))^2\sin^2(u)\sin^2(v))\Big)}\Big(\tfrac{1}{(a-c\cos(u)\cos(v))^4}(2b^2c^2\cos(u)\sin(u)\sin^2(v)(a-d\cos(v))^2-$$
$$2ab^2(a\cos(u)-c\cos(v))\sin(u)(a-d\cos(v))^2+2\cos(u)(ab^2-(a^2-c^2)d\cos(v))^2\sin(u))-\tfrac{1}{(a-c\cos(u)\cos(v))^5}$$
$$4c\cos(v)\sin(u)(b^2(a\cos(u)-c\cos(v))^2(a-d\cos(v))^2+b^2c^2\sin^2(u)\sin^2(v)(a-d\cos(v))^2+(ab^2-(a^2-c^2)d\cos(v))^2\sin^2(u))\Big)\bigg/$$
$$\Big(\tfrac{1}{(a-c\cos(u)\cos(v))^4}(b^2(a\cos(u)-c\cos(v))^2(a-d\cos(v))^2+b^2c^2\sin^2(u)\sin^2(v)(a-d\cos(v))^2+(ab^2-(a^2-c^2)d\cos(v))^2\sin^2(u))\Big)^{3/2}+$$
$$\Big(5\Big(\tfrac{1}{(a-c\cos(u)\cos(v))^4}(2a^2b^2c(d-c\cos(u))\sin^2(v)\sin^3(u)+2b^2c(d-c\cos(u))(c\cos(u)-a\cos(v))^2\sin(u)+$$
$$2a^2b^2\cos(u)(d-c\cos(u))^2\sin^2(v)\sin(u)-2\cos(u)(c\cos(u)b^2+(c^2-a^2)d)^2\sin^2(v)\sin(u)-2b^2c\cos^2(u)$$
$$(c\cos(u)b^2+(c^2-a^2)d)\sin^2(v)\sin(u)-2b^2c(d-c\cos(u))^2(c\cos(u)-a\cos(v))\sin(u))-\tfrac{1}{(a-c\cos(u)\cos(v))^5}4c\cos(v)\sin(u)$$
$$(b^2(d-c\cos(u))^2(c\cos(u)-a\cos(v))^2+\cos^2(u)(c\cos(u)b^2+(c^2-a^2)d)^2\sin^2(v)+a^2b^2(d-c\cos(u))^2\sin^2(u)\sin^2(v))\Big)\bigg/$$
$$\Big(\sqrt{\Big(\tfrac{1}{(a-c\cos(u)\cos(v))^4}(b^2(d-c\cos(u))^2(c\cos(u)-a\cos(v))^2+\cos^2(u)(c\cos(u)b^2+(c^2-a^2)d)^2\sin^2(v)+}$$
$$\overline{a^2b^2(d-c\cos(u))^2\sin^2(u)\sin^2(v))\Big)}$$
$$\sqrt{\Big(\tfrac{1}{(a-c\cos(u)\cos(v))^4}(b^2(a\cos(u)-c\cos(v))^2(a-d\cos(v))^2+b^2c^2\sin^2(u)\sin^2(v)(a-d\cos(v))^2+(ab^2-(a^2-c^2)d\cos(v))^2\sin^2(u))\Big)}-$$
$$\Big(\tfrac{1}{(a-c\cos(u)\cos(v))^4}(2b^2c^2d(a-d\cos(v))\sin^2(u)\sin^3(v)+2b^2c(a\cos(u)-c\cos(v))(a-d\cos(v))^2\sin(v)+2b^2c^2\cos(v)(a-d\cos(v))^2\sin^2(u)$$
$$\sin(v)+2(a^2-c^2)d(ab^2-(a^2-c^2)d\cos(v))\sin^2(u)\sin(v)+2b^2d(a\cos(u)-c\cos(v))^2(a-d\cos(v))\sin(v))-\tfrac{1}{(a-c\cos(u)\cos(v))^5}$$
$$4c\cos(u)\sin(v)(b^2(a\cos(u)-c\cos(v))^2(a-d\cos(v))^2+b^2c^2\sin^2(u)\sin^2(v)(a-d\cos(v))^2+(ab^2-(a^2-c^2)d\cos(v))^2\sin^2(u))\Big)\bigg/$$
$$\Big(\sqrt{\Big(\tfrac{1}{(a-c\cos(u)\cos(v))^4}(b^2(d-c\cos(u))^2(c\cos(u)-a\cos(v))^2+\cos^2(u)(c\cos(u)b^2+(c^2-a^2)d)^2\sin^2(v)+}$$
$$\overline{a^2b^2(d-c\cos(u))^2\sin^2(u)\sin^2(v))\Big)}\sqrt{\Big(\tfrac{1}{(a-c\cos(u)\cos(v))^4}}$$
$$\overline{(b^2(a\cos(u)-c\cos(v))^2(a-d\cos(v))^2+b^2c^2\sin^2(u)\sin^2(v)(a-d\cos(v))^2+(ab^2-(a^2-c^2)d\cos(v))^2\sin^2(u))\Big)}\Big)\bigg/$$
$$\Big(\sqrt{\Big(\tfrac{1}{(a-c\cos(u)\cos(v))^4}(b^2(d-c\cos(u))^2(c\cos(u)-a\cos(v))^2+\cos^2(u)(c\cos(u)b^2+(c^2-a^2)d)^2\sin^2(v)+a^2b^2(d-c\cos(u))^2\sin^2(u)\sin^2(v))\Big)}$$
$$\sqrt{\Big(\tfrac{1}{(a-c\cos(u)\cos(v))^4}}$$
$$\overline{(b^2(a\cos(u)-c\cos(v))^2(a-d\cos(v))^2+b^2c^2\sin^2(u)\sin^2(v)(a-d\cos(v))^2+(ab^2-(a^2-c^2)d\cos(v))^2\sin^2(u))\Big)\Big)}$$

Eq. (16)

The following is the expression for the solution for $\alpha$ in Eq. (13):

$\alpha =$

Out[291]//TraditionalForm=

$$-\Bigg(\Bigg(\Bigg(\sqrt{\Big(\frac{1}{(a-c\cos(u)\cos(v))^4}(b^2(a\cos(u)-c\cos(v))^2(a-d\cos(v))^2+b^2c^2\sin^2(u)\sin^2(v)(a-d\cos(v))^2+(ab^2-(a^2-c^2)d\cos(v))^2\sin^2(u))\Big)}$$

$$\Big(\frac{1}{(a-c\cos(u)\cos(v))^4}(2a^2b^2\cos(v)\sin^2(u)\sin(v)(d-c\cos(u))^2+2ab^2(c\cos(u)-a\cos(v))\sin(v)(d-c\cos(u))^2+$$

$$2\cos^2(u)(c\cos(u)b^2+(c^2-a^2)d)^2\cos(v)\sin(v))-\frac{1}{(a-c\cos(u)\cos(v))^5}$$

$$4c\cos(u)\sin(v)(b^2(d-c\cos(u))^2(c\cos(u)-a\cos(v))^2+\cos^2(u)(c\cos(u)b^2+(c^2-a^2)d)^2\sin^2(v)+a^2b^2(d-c\cos(u))^2\sin^2(u)\sin^2(v))\Bigg)\Bigg/$$

$$\Big(\frac{1}{(a-c\cos(u)\cos(v))^4}(b^2(d-c\cos(u))^2(c\cos(u)-a\cos(v))^2+\cos^2(u)(c\cos(u)b^2+(c^2-a^2)d)^2\sin^2(v)+a^2b^2(d-c\cos(u))^2\sin^2(u)\sin^2(v))\Big)^{3/2}-$$

$$\Bigg(5\sqrt{\Big(\frac{1}{(a-c\cos(u)\cos(v))^4}(b^2(d-c\cos(u))^2(c\cos(u)-a\cos(v))^2+\cos^2(u)(c\cos(u)b^2+(c^2-a^2)d)^2\sin^2(v)+a^2b^2(d-c\cos(u))^2\sin^2(u)\sin^2(v))\Big)}$$

$$\Big(\frac{1}{(a-c\cos(u)\cos(v))^4}(2b^2c^2\cos(u)\sin(u)\sin^2(v)(a-d\cos(v))^2-2ab^2(a\cos(u)-c\cos(v))\sin(u)(a-d\cos(v))^2+$$

$$2\cos(u)(ab^2-(a^2-c^2)d\cos(v))^2\sin(u))-\frac{1}{(a-c\cos(u)\cos(v))^5}$$

$$4c\cos(v)\sin(u)(b^2(a\cos(u)-c\cos(v))^2(a-d\cos(v))^2+b^2c^2\sin^2(u)\sin^2(v)(a-d\cos(v))^2+(ab^2-(a^2-c^2)d\cos(v))^2\sin^2(u))\Bigg)\Bigg/$$

$$\Big(\frac{1}{(a-c\cos(u)\cos(v))^4}(b^2(a\cos(u)-c\cos(v))^2(a-d\cos(v))^2+b^2c^2\sin^2(u)\sin^2(v)(a-d\cos(v))^2+(ab^2-(a^2-c^2)d\cos(v))^2\sin^2(u))\Big)^{3/2}+$$

$$\Bigg(5\Big(\frac{1}{(a-c\cos(u)\cos(v))^4}(2a^2b^2c(d-c\cos(u))\sin^2(v)\sin^3(u)+2b^2c(d-c\cos(u))(c\cos(u)-a\cos(v))^2\sin(u)+$$

$$2a^2b^2\cos(u)(d-c\cos(u))^2\sin^2(v)\sin(u)-2\cos(u)(c\cos(u)b^2+(c^2-a^2)d)^2\sin^2(v)\sin(u)-$$

$$2b^2c\cos^2(u)(c\cos(u)b^2+(c^2-a^2)d)\sin^2(v)\sin(u)-2b^2c(d-c\cos(u))^2(c\cos(u)-a\cos(v))\sin(u))-\frac{1}{(a-c\cos(u)\cos(v))^5}$$

$$4c\cos(v)\sin(u)(b^2(d-c\cos(u))^2(c\cos(u)-a\cos(v))^2+\cos^2(u)(c\cos(u)b^2+(c^2-a^2)d)^2\sin^2(v)+a^2b^2(d-c\cos(u))^2\sin^2(u)\sin^2(v))\Big)\Bigg/$$

$$\Bigg(\sqrt{\Big(\frac{1}{(a-c\cos(u)\cos(v))^4}(b^2(d-c\cos(u))^2(c\cos(u)-a\cos(v))^2+\cos^2(u)(c\cos(u)b^2+(c^2-a^2)d)^2\sin^2(v)+a^2b^2(d-c\cos(u))^2\sin^2(u)\sin^2(v))\Big)}$$

$$\sqrt{\Big(\frac{1}{(a-c\cos(u)\cos(v))^4}(b^2(a\cos(u)-c\cos(v))^2(a-d\cos(v))^2+b^2c^2\sin^2(u)\sin^2(v)(a-d\cos(v))^2+(ab^2-(a^2-c^2)d\cos(v))^2\sin^2(u))\Big)}\Bigg)-$$

$$\Big(\frac{1}{(a-c\cos(u)\cos(v))^4}(2b^2c^2d(a-d\cos(v))\sin^2(u)\sin^3(v)+2b^2c(a\cos(u)-c\cos(v))(a-d\cos(v))^2\sin(v)+2b^2c^2\cos(v)(a-d\cos(v))^2$$

$$\sin^2(u)\sin(v)+2(a^2-c^2)d(ab^2-(a^2-c^2)d\cos(v))\sin^2(u)\sin(v)+2b^2d(a\cos(u)-c\cos(v))^2(a-d\cos(v))\sin(v))-\frac{1}{(a-c\cos(u)\cos(v))^5}$$

$$4c\cos(u)\sin(v)(b^2(a\cos(u)-c\cos(v))^2(a-d\cos(v))^2+b^2c^2\sin^2(u)\sin^2(v)(a-d\cos(v))^2+(ab^2-(a^2-c^2)d\cos(v))^2\sin^2(u))\Big)\Bigg/$$

$$\Bigg(\sqrt{\Big(\frac{1}{(a-c\cos(u)\cos(v))^4}(b^2(d-c\cos(u))^2(c\cos(u)-a\cos(v))^2+\cos^2(u)(c\cos(u)b^2+(c^2-a^2)d)^2\sin^2(v)+a^2b^2(d-c\cos(u))^2\sin^2(u)\sin^2(v))\Big)}$$

$$\sqrt{\Big(\frac{1}{(a-c\cos(u)\cos(v))^4}(b^2(a\cos(u)-c\cos(v))^2(a-d\cos(v))^2+b^2c^2\sin^2(u)\sin^2(v)(a-d\cos(v))^2+(ab^2-(a^2-c^2)d\cos(v))^2\sin^2(u))\Big)}\Bigg)\Bigg/$$

$$\Bigg(\sqrt{\Big(\frac{1}{(a-c\cos(u)\cos(v))^4}(b^2(d-c\cos(u))^2(c\cos(u)-a\cos(v))^2+\cos^2(u)(c\cos(u)b^2+(c^2-a^2)d)^2\sin^2(v)+a^2b^2(d-c\cos(u))^2\sin^2(u)\sin^2(v))\Big)}$$

$$\sqrt{\Big(\frac{1}{(a-c\cos(u)\cos(v))^4}(b^2(a\cos(u)-c\cos(v))^2(a-d\cos(v))^2+b^2c^2\sin^2(u)\sin^2(v)(a-d\cos(v))^2+(ab^2-(a^2-c^2)d\cos(v))^2\sin^2(u))\Big)}$$

$$\Big(-\big(10/\big(\sqrt{((b^2(a\cos(u)-c\cos(v))^2(a-d\cos(v))^2+b^2c^2\sin^2(u)\sin^2(v)(a-d\cos(v))^2+(ab^2-(a^2-c^2)d\cos(v))^2\sin^2(u))/(a-c\cos(u)\cos(v))^4)}\big)\big)-$$

$$2/\big(\sqrt{((b^2(d-c\cos(u))^2(c\cos(u)-a\cos(v))^2+\cos^2(u)(c\cos(u)b^2+(c^2-a^2)d)^2\sin^2(v)+a^2b^2(d-c\cos(u))^2\sin^2(u)\sin^2(v))/}$$

$$(a-c\cos(u)\cos(v))^4)}\big)\Big)\Bigg)\Bigg)$$

Eq. (17)